\documentclass[prx,aps,twocolumn,reprint,superscriptaddress,showpacs,floatfix,longbibliography,footinbib]{revtex4-1}
\usepackage{mathtools}
\usepackage{amsfonts}
\DeclarePairedDelimiter\bra{\langle}{\rvert}
\DeclarePairedDelimiter\ket{\lvert}{\rangle}

\begin{document}

\title{A long-lived spin-orbit-coupled degenerate dipolar {Fermi} gas}

\author{Nathaniel Q. Burdick}
\affiliation{Department of Applied Physics, Stanford University, Stanford CA 94305}
\affiliation{E. L. Ginzton Laboratory, Stanford University, Stanford CA 94305}
\author{Yijun Tang}
\affiliation{E. L. Ginzton Laboratory, Stanford University, Stanford CA 94305}
\affiliation{Department of Physics, Stanford University, Stanford CA 94305}
\author{Benjamin L. Lev}
\affiliation{Department of Applied Physics, Stanford University, Stanford CA 94305}
\affiliation{E. L. Ginzton Laboratory, Stanford University, Stanford CA 94305}
\affiliation{Department of Physics, Stanford University, Stanford CA 94305}

\date{\today}

\begin{abstract}
 We describe the creation of a long-lived spin-orbit-coupled gas of quantum degenerate atoms using the most magnetic fermionic element, dysprosium. Spin-orbit-coupling arises from a synthetic gauge field created by the adiabatic following of degenerate dressed states comprised of optically coupled components of an atomic spin. Because of dysprosium's large electronic orbital angular momentum and large magnetic moment, the lifetime of the gas is limited not by spontaneous emission from the light-matter coupling, as for gases of alkali-metal atoms, but by dipolar relaxation of the spin.  This relaxation is  suppressed at large magnetic fields due to Fermi statistics.   We observe lifetimes up to 400~ms, which exceeds that of spin-orbit-coupled fermionic alkali atoms by a factor of 10--100, and is close to the value obtained from a theoretical model.   Elastic dipolar interactions are also observed to influence the Rabi evolution of the spin, revealing an interacting fermionic system.  The long lifetime of this weakly interacting spin-orbit-coupled degenerate Fermi gas will facilitate the study of quantum many-body phenomena manifest at longer timescales, with exciting implications for the exploration of exotic topological quantum liquids.  
\end{abstract}

\maketitle

New classes of topologically nontrivial materials require the coupling of charge carriers' quantum-mechanical spin to their momentum. This spin-orbit coupling (SOC) arises from electron movement through the intrinsic electric field of the crystal and can result in topological insulators and superconductors as well as exotic quantum Hall states~\cite{Hasan2010,*Qi2011}.   Ultracold neutral  atoms can experience an analogous coupling in the presence of light fields that couple Zeeman sublevels via a two-photon Raman transition~\cite{Lin2011}.  Recoil momentum is transferred as the optically coupled spin flips~\cite{Higbie:2002gb,Liu:2009hj,Spielman2009}.   The adiabatic evolution of these dressed states as the atom moves in the Raman field creates a synthetic gauge field.  This field takes the form of spin-orbit coupling when these states form a degenerate manifold~\cite{Dalibard:2011gg}. Spin-orbit coupling has been created in this manner in Bose-Einstein condensates (BECs)~\cite{Lin2011,Williams2011,Zhang:2012fd,Qu:2013bf,Hamner2014,Olson2014,Ji:2014jh,Ji:2015ih,Jimenez-Garcia2015} and  degenerate Fermi gases (DFGs)~\cite{Cheuk2012,Wang2012,Huang2016,Williams2013,Fu:2014bq} of alkali-metals.  

These achievements open new avenues to experimentally study topological matter not realizable in the solid state~\cite{Galitski:2013dh,Zhang:2014eh,*Goldman:2014bva,*Zhai:2015hg}. Specifically, novel topological superfluids and other quantum liquids may be observable in long-lived, interacting spin-orbit-coupled Fermi gases~\cite{Galitski:2013dh,Sato:2009id,,*Jiang:2011cw,*Zhu:2011cb,*Alicea:2012hz,*Wei:2012di,*Liu:2012dl,*Cui:2014dp,*Ruhman:2015dk,Nascimbene:2013fr,Celi:2014dg}. Such investigations would benefit from the well-characterized Hamiltonians, controllable interactions and disorder, and  tunable dimensionality inherent to ultracold atomic systems.  However, a consequence of employing Raman coupling to generate  SOC with alkali atoms is atomic heating due to spontaneous emission.  This heating leads to loss of quantum degeneracy and trap population and severely limits the lifetimes of fermionic alkali gases~\cite{Cheuk2012,Wang2012,Huang2016}, hampering the  study of quantum many-body phenomena manifest at longer timescales.

Synthetic gauge fields have been created in gases of fermionic alkaline-earth atoms~\cite{Wall:2016cl,Mancini:2015fba} using narrow optical transitions, though under the condition of optical lattice confinement.  By using atoms with ground state orbital angular momentum $L>0$, spontaneous emission can be eliminated while still producing large Raman coupling even without lattice confinement.  This provides more flexibility for investigating a wide range of quantum systems. Specifically, large $L$'s ensure that the vector and tensor polarizabilities which give rise to Raman coupling matrix elements always scale as inverse atomic detuning $\Delta_a^{-1}$~\cite{Deutsch:1998jn,*Geremia:2006dv}.  Since spontaneous emission decreases  rapidly with detuning as $\Delta_a^{-2}$, one can always choose a detuning that provides a large Raman coupling $\Omega_{R}$ while minimizing heating from incoherent scattering~\cite{Cui2013}.

The open-shell lanthanide atoms Dy~\cite{Lu2011a,Lu2012} and Er~\cite{Aikawa2012,Aikawa2014} are suitable candidates with $L=6$ and $L=5$, respectively. Moreover, these atoms' large total spin---e.g., $F=21/2$ for $^{161}$Dy~\footnote{The ground state hyperfine manifold is spin $F=J+I=21/2$ for $^{161}$Dy, where the total electronic angular momentum is $J=L+S=8$, the total electronic spin is $S = 2$, and the nuclear spin is $I=5/2$.}---can enable the study of  (possibly non-Abelian) spinors coupled to large synthetic gauge fields~\cite{Cui2013,Ho:2015hf}.  
 Though spontaneous emission can be eliminated, the strong dipole-dipole interaction (DDI) between these highly magnetic, large-spin atoms induces decay from metastable Zeeman levels that leads to heating and trap loss~\cite{Hensler2003,Pasquiou2010,Burdick2015}.  Metastable states are  unavoidable when using a Raman coupling scheme to produce SOC. However, by taking advantage of the suppressed relaxation rates of collisions between identical fermions~\cite{Burdick2015}, we are able to produce SOC gases of $^{161}$Dy with lifetimes as long as 400~ms, surpassing that of $^{40}$K and $^{6}$Li fermionic SOC systems by  factors of $\sim$10 and 100, respectively~\cite{Cheuk2012,Wang2012,Huang2016}, and comparable to the lifetime of bosonic spin-orbit coupled alkali gases~\cite{Lin2011}.

  \begin{figure}[t!]
    \includegraphics[width=\columnwidth]{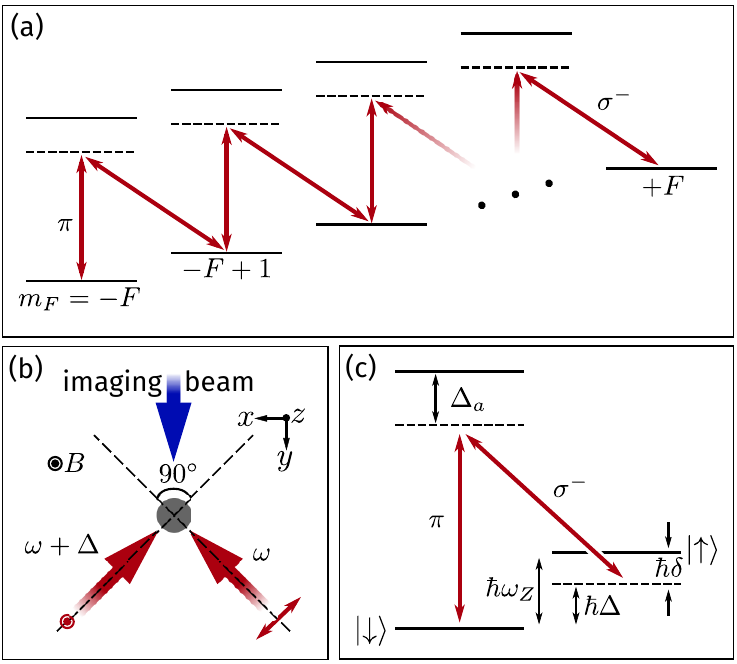}
    \caption{(a) Raman coupling with $\pi$ and $\sigma^{-}$ transitions for a large $F$ atom with zero or small quadratic Zeeman shift. (b) Schematic of Raman beam geometry used to produce SOC with $^{161}$Dy. (c) Coupling scheme for $^{161}$Dy at a magnetic field high enough that  with the lowest two Zeeman states are isolated by the quadratic Zeeman shift. The states that form this pseudospin-1/2 system are labeled $\ket{\downarrow}=\ket{m_{F}=-21/2}$ and $\ket{\uparrow}=\ket{m_{F}=-19/2}$.}
    \label{couplingscheme}
  \end{figure}

\textit{Raman-coupling field configuration.}
We measure the effects of both elastic and inelastic dipolar collisions on gases of fermionic $^{161}$Dy under SOC using the Raman coupling scheme sketched in Fig.~\ref{couplingscheme}(b)-(c).  The lowest two Zeeman sublevels of the lowest ground state hyperfine manifold, $\ket*{\downarrow}\equiv\ket*{F=21/2,m_{F}=-21/2}$ and $\ket*{\uparrow}\equiv\ket*{21/2,-19/2}$, are coupled by the two Raman lasers with wavelength $\lambda=741$~nm and coupling strength $\Omega_{R}$. The Raman lasers are derived from a single Ti-sapphire laser that is detuned by several GHz from the 1.78(2)-kHz-wide 741-nm transition in $^{161}$Dy \cite{Lu2011}.
The first Raman beam propagates along $(\hat{x}-\hat{y})$ with frequency $\omega$ and is linearly polarized along $\hat{x}+\hat{y}$; i.e., it drives $\sigma^{\pm}$ transitions, though $\sigma^{+}$ are off resonant and not shown in Fig.~\ref{couplingscheme}.  The second Raman beam propagates along $-(\hat{x}+\hat{y})$ with frequency $\omega+\Delta$ and is linearly polarized along $\hat{z}$; i.e., it drives $\pi$ transitions (the magnetic field is along $\hat{z}$). Each beam is frequency shifted and controlled by an acousto-optical modulator (AOM) with two channels of a single frequency generator driving the two AOMs in order to maintain phase coherence. The detuning from two-photon Raman resonance is defined as $\delta = \omega_{Z}-\Delta$, where $\hbar\omega_Z$ is the Zeeman energy. Each Raman transition transfers $2\hbar k_{R}$ of momentum between an atom and the lasers, where $k_{R}=k_{r}\sin\left(\theta/2\right)$, $k_{r}=2\pi/\lambda$ is the single-photon recoil momentum, and $\theta=90^{\circ}$ is the  angle between beams. The natural unit of energy is $E_{R} = (\hbar k_{R})^{2}/2m$, where $m$ is the mass. For our configuration, $E_{R}/h = 1.1(1)$~kHz, with the uncertainty from beam alignment angle.

 To suppress dipolar relaxation, we apply a homogeneous bias field of $B_0 = 33.846(5)$~G  along $\hat{z}$, producing a Zeeman splitting $\hbar\omega_{Z} = h \times 43.986(6)$~MHz, varying as  1.275~MHz/G at this field value~\footnote{All reported errors are standard errors unless otherwise noted.}. Moreover,  the quadratic Zeeman shift at this field is sufficient to produce a two-photon detuning between $\ket{m_{F}=-19/2}$ and $\ket{m_{F}=-17/2}$ of more than $70E_{R}/\hbar$.  This allows us to neglect all Raman coupling to spin states with $m_{F}>-19/2$ when the Raman fields are on resonance with  $\ket{m_{F}=-21/2}$ and $\ket{m_{F}=-19/2}$, i.e., when $\delta=0$. Indeed, we  observe (via Stern-Gerlach measurement) population in only these two   states.  Restricting to a pseudospin-1/2 system facilitates  fermionic suppression of dipolar relaxation~\cite{Burdick2015}.

  \begin{figure}[t!]
    \includegraphics[width=\columnwidth]{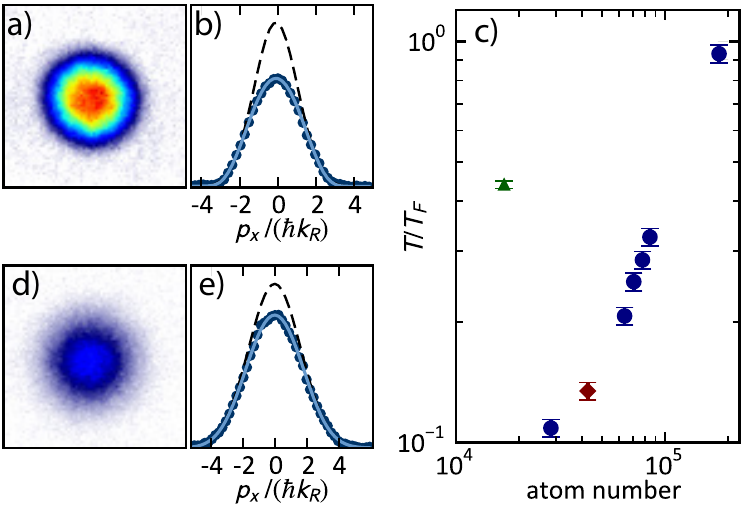}
    \caption{
      (a) DFG of $^{161}$Dy at 0.48(1)~G with $N=4.3(2)\times10^{4}$ atoms and $T/T_{F}=0.13(1)$.
      (b) 1D integrated momentum distribution of DFG in (a) with Thomas-Fermi fit (solid line) and thermal fit to the wings (dashed line).
      (c) Single-species evaporation trajectory of $^{161}$Dy. The red diamond corresponds to the DFG shown in (a). The green triangle corresponds to the DFG shown in (d).
      (d) DFG of $^{161}$Dy at $B_0=33.846(5)$ with $N=1.7(3)\times10^{4}$ and $T/T_{F}=0.44(1)$.
      (e) Same as in (b) but for DFG in (d).
    }
    \label{dfgfig}
  \end{figure}

\textit{Degenerate Fermi gas preparation.} We prepare ultracold gases of fermionic $^{161}$Dy via a 421-nm magneto-optical trap (MOT) followed by a 741-nm MOT as in Ref.~\cite{Lu2012}. We transfer ${\sim}3\times10^{6}$ atoms to a 1064-nm optical dipole trap and perform rf adiabatic rapid passage (ARP) to prepare the atoms in the absolute ground state, $\ket{F=21/2,m_{F}=-21/2}$. A DFG is produced via forced evaporation in a crossed optical dipole trap formed by two additional 1064-nm beams at low field $0.48(1)$~G. The elastic DDI mediates collisions between identical fermions even below the $p$-wave threshold, allowing for single species thermalization and evaporation down to degeneracy \cite{Lu2012,Aikawa2014}. Figures~\ref{dfgfig}(a) and~\ref{dfgfig}(b) show an absorption image of a $^{161}$Dy DFG at $0.48(1)$~G  with $N= 4.3(2)\times10^{4}$ atoms and a ratio of temperature to Fermi temperature of $T/T_{F}=0.13(1)$.   The spin-polarized evaporation trajectory is shown in Fig.~\ref{dfgfig}(c).  The dense Feshbach resonance spectrum of $^{161}$Dy results in atom loss when sweeping to higher magnetic field~\cite{Baumann2014}. Nevertheless, a relatively high-field ($\sim$34~G) DFG of $1.7(3)\times10^{4}$ atoms at $T/T_{F}=0.44(1)$ is shown in Fig.~\ref{dfgfig}(c)-(d). The increase in $T/T_{F}$ compared to the low-field DFG  is both due to atom loss ($T_{F}\propto N^{1/3}$) and to heating from anti-evaporation~\footnote{Colder, denser atoms are more likely to be removed when sweeping to high field.}~\cite{Grimm03}.

\begin{figure*}[t!]
  \includegraphics[width=\textwidth]{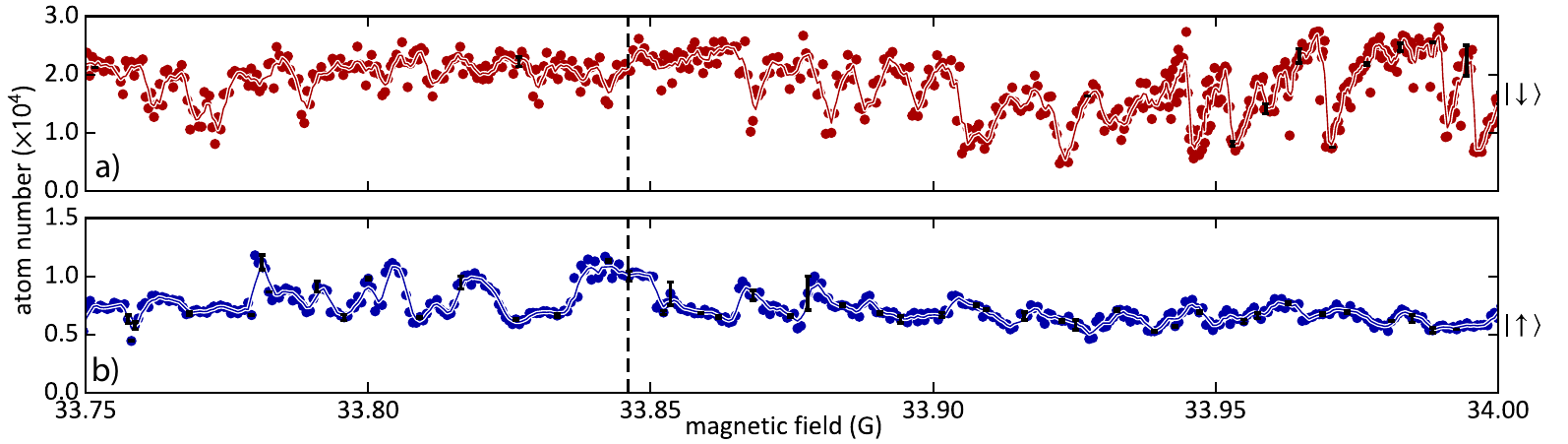}
  \caption{Atom loss spectroscopy of Feshbach resonances for $^{161}$Dy in (a) $\ket{\downarrow}\equiv\ket{m_{F}=-21/2}$; (b) $\ket{\uparrow}\equiv\ket{m_{F}=-19/2}$.   Both spectra were taken with the same initial atom number; note the different $y$-axis scales. The vertical dashed line indicates the field used to produce SOC fields, $B_0=33.846(5)$~G, and the solid lines are guides to the eye. The black error bars represent statistical errors on the subset of points indicated.}
  \label{feshbach}
\end{figure*}

\textit{Ultradense Feshbach-resonance spectra.} The dense Feshbach resonance landscape of lanthanide atoms \cite{Aikawa2012,Baumann2014,Frisch2014,Maier2015,Maier2015a} requires special consideration when choosing a magnetic field at which to implement SOC.  We find that this density---already large at low field~\cite{Baumann2014}---significantly increases at higher fields.  Atom  spectroscopy for $^{161}$Dy near 34~G over a 250-mG window is shown in Fig.~\ref{feshbach} for spin-polarized gases of $\ket{\downarrow}$ and $\ket{\uparrow}$. 
We lower the density of the gas when sweeping to high field in order to reduce collisional loss during the sweep itself. To do so, we reduce the  trap frequency from $\bar{f}=150(10)$~Hz (geometric mean) to ${\sim}90$~Hz in 100~ms. We then jump the field to the desired field. For the data in Fig.~\ref{feshbach}(a), we the wait 100~ms while the eddy currents in the chamber damp.  During this period we recompress the gas and then hold the gas for an additional 60~ms in the fully compressed trap before measuring atom number.  We use as similar a procedure as possible for the data in Fig.~\ref{feshbach}(b), even though a spin-flip is required. Specifically,  we first  jump the field to 33.967(5)~G, wait 100~ms for eddy currents to damp, and then apply an rf ARP pulse to the gas in 15~ms to flip the spin to  $\ket{\uparrow}$.  We measure no loss during the 100~ms before ARP because 33.967(5)~G is a resonance-free field for $\ket{\downarrow}$ and the gas is at low density. The field is then jumped to its final value before the trap is recompressed in 100~ms.  Finally, the gas is held for an additional 60~ms, as for the  $\ket{\downarrow}$ data.

We find that the density of loss features in the  doubly spin-polarized $|F=J+I|=|m_F|$ spectrum of the  $\ket{\downarrow}$ state is $\sim$10$\times$ greater at this higher field than at low field between 0--6~G~\cite{Baumann2014}.  This is perhaps due to a larger Zeeman-induced coupling among the molecular potentials~\cite{Maier2015}.  Resonance-free regions are sparse and only a few tens of mG wide. The loss spectrum for the non-doubly spin-polarized state $\ket{\uparrow}$ has even fewer favorable regions with each  only a few mG wide.  Indeed, these ultradense Feshbach-resonance spectra  might not even support completely resonance-free regions.  This greater relative loss might be due to a higher density of resonances in this entrance channel~\footnote{J.~Bohn, private communication (2016).}.  We note that overlapping resonances are known from nuclear physics to lead to Ericson fluctuations in resonance spectra similar to that observed here~\cite{Mayle2012}.  Future work will consider the possible role of Ericson fluctuations in $^{161}$Dy loss spectra.

The magnetic field $B_0$ is chosen to both  minimize three-body loss for  spin-polarized gases in $\ket{\downarrow}$ and $\ket{\uparrow}$ states separately and to minimize loss in superpositions of these states:  We verified through a measurement of the atom loss spectrum of mixed $\ket{\downarrow}$ and $\ket{\uparrow}$ states that this field sits in a region of near-minimal loss.  We also observed that  the features in Fig.~\ref{feshbach}(a) exhibit no dependence on coupling strength, unlike that found  for alkali-atom gases under Raman coupling~\cite{Williams2011,Williams2013,Fu:2014bq}.   In those experiments, low-order single partial-wave Feshbach resonances were  modified by SOC to induce higher-order effective partial-wave contributions.  In Dy, however, many partial waves already contribute to Feshbach resonances~\cite{Petrov2012}, possibly rendering the SOC modification negligible.

\textit{Spin-orbit coupling measurement.} 
The Raman coupling is described by the effective single-particle Hamiltonian
\begin{equation}
\begin{aligned}
\hat{H}_{R} &=  \frac{\left[\hat{p}_{x}-2\hbar k_{R}(\hat{F}_{z}+F)\right]^{2}}{2m} 
- \hbar\delta \hat{F}_{z} - \frac{\hbar\Omega_{R}}{2}\left(\hat{\Lambda} + \hat{\Lambda}^{\dagger} \right),\\
\hat{\Lambda} &= \hat{F}_{+}\left(\hat{\mathbb{I}}-\frac{2\hat{F}_{z}+\hat{\mathbb{I}}}{2F+3}\right)\left[\frac{2F+3}{\sqrt{2F}(4F+2)}\right],
\label{SOH1}
\end{aligned}
\end{equation}
for the bare atomic states $\ket{\downarrow,p_{x}}$ and $\ket{\uparrow,p_{x}+2\hbar k_{R}}$, where $\hat{F}_{z}$ is the spin-projection operator, $\hat{F}_{+}$ is the raising operator, $\hat{\mathbb{I}}$ is the identity, and $\hat{p}_{x}$ is the quasimomentum. This Hamiltonian contains equal strengths of Rashba and Dresselhaus SOC \cite{Cui2013,Bychkov1984,*Dresselhaus1955}.

\begin{figure}[t!]
  \includegraphics[width=\columnwidth]{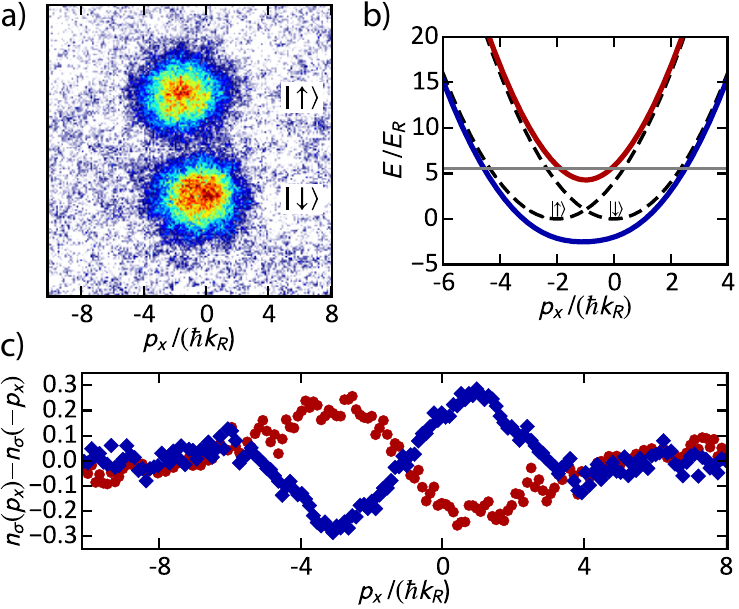}
  \caption{
    (a) Momentum distribution of a SOC-gas of $^{161}$Dy with $\Omega_{R}=5.9(2)E_{R}$, $\delta=0.1(2)E_{R}$, and $T/T_{F}=0.4^{+0.2}_{-0.1}$ after diabatically removing the coupling and separating the spin states with a magnetic field gradient.  Time-of-flight duration is 10~ms.
    (b) SOC dispersion curve for the gas in panel (a). The solid horizontal line indicates the Fermi energy for a gas of $N=1.2\times10^{4}$ atoms in a trap with $\bar{f}=150\ \mathrm{Hz}$.
    (c) Integrated momentum asymmetry along $x$ for $\sigma=\ \downarrow$ (blue diamonds) and $\sigma=\ \uparrow$ (red circles).}
  \label{socfig}
  \end{figure}

The gas is loaded into the lowest SOC-dressed band by adiabatically turning on the Raman coupling over a time period of 35~ms. We hold the gas in this configuration for a time $\tau_{soc}$. By then diabatically turning off the coupling lasers, the quasimomentum of the dressed band is converted into mechanical momentum, which can be measured via time-of-flight imaging \cite{Lin2009,Fu2011}. A gas with $\Omega_{R} = 5.9(2)E_{R}$ is shown in Fig.~\ref{socfig}(a) with the spin states separated by a magnetic field gradient. The corresponding quasimomentum dispersion is shown in Fig.~\ref{socfig}(b). The SOC manifests itself as a momentum asymmetry in the bare spin states, i.e., $n_{\sigma}(p_{x})\neq n_{\sigma}(-p_{x})$ for $\sigma=\{\downarrow,\uparrow\}$ \cite{Wang2012}. This momentum asymmetry is highlighted in Fig.~\ref{socfig}(c) by plotting the reflected momentum difference of the two spin states, $n_{\sigma}(p_{x})-n_{\sigma}(-p_{x})$.
  \begin{figure}[th!]
    \includegraphics[width=1.\columnwidth]{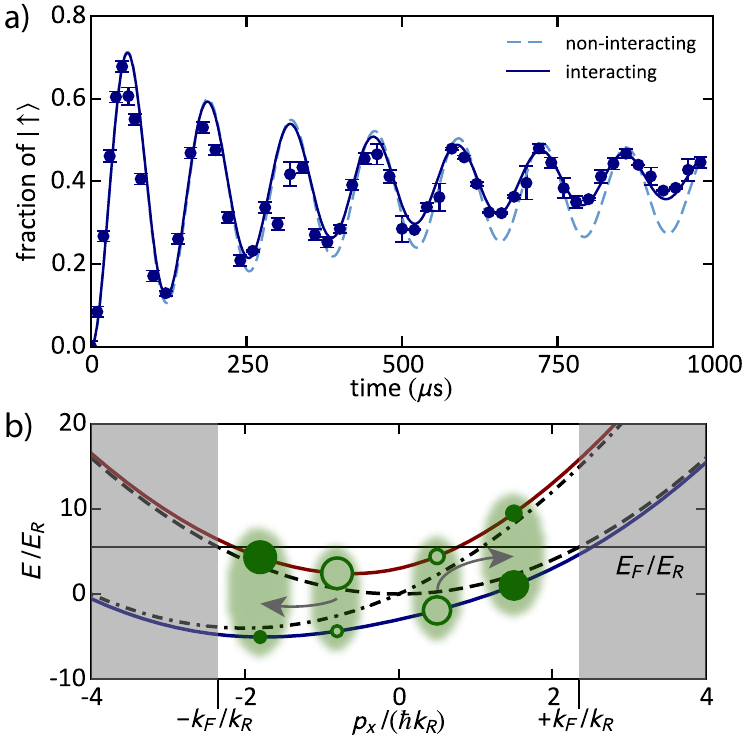}
    \caption{(a) Rabi oscillation data with $\Omega_{R}=6.8(2)E_{R}$ and $\delta=-3.8(2)E_{R}$. Predicted spin dynamics for a non-interacting Fermi gas (dashed line) according to Eq.~\ref{osc_eq} and an interacting Fermi gas (solid line) according to a Monte-Carlo simulation.  (b) Cartoon illustrating elastic collisions during Rabi oscillations.  The dashed (dot-dashed) line is the uncoupled $\ket{\downarrow}$ ($\ket{\uparrow}$) quasimomentum dispersion at $B_0$. The solid colored quasimomentum dispersions are for the dressed states with $\Omega_{R}=6E_{R}$ and $\delta=-4E_{R}$. The Fermi energy is marked by the solid horizontal line, and an uncoupled $\ket{\downarrow}$-polarized Fermi gas at $T=0$ occupies the quasimomenta between $\pm k_{F}$ (unshaded region). See text for explanation.}
  \label{rabifig}
  \end{figure}

We are currently unable to perform spin-injection or momentum-resolved radio-frequency spectroscopy, as has been done in the fermionic SOC $^{6}$Li and $^{40}$K systems~\cite{Cheuk2012,Wang2012}.  The relatively large mass of $^{161}$Dy means that to achieve a similar band structure resolution, field stability would have to be less than  a few hundreds of~$\mu$G,  a factor of ten lower than our current capability.

\textit{Elastic dipolar collisions in Rabi oscillations.} In order to calibrate the strength of the Raman coupling, we drive Rabi oscillations by diabatically turning on the coupling lasers in $<$5~$\mu$s. For a non-interacting system, the probability of observing an atom in state $\ket{\uparrow}$ after being initialized in state $\ket{\downarrow}$ is \cite{Wang2012}
\begin{equation}
P_{\uparrow}(k_{x},t) =
\frac{  \sin^{2}  \left(  \sqrt{\big(\frac{\hbar k_{x}k_{R}}{m} + \frac{\delta}{2}\big)^{2} +    \big(\frac{\Omega_{R}}{2}\big)^{2}}  t  \right)}{  1 + \left(\frac{2\hbar k_{x}k_{R}}{\Omega_{R}m}+\frac{\delta}{\Omega_{R}}\right)^{2}},
\label{osc_eq}
\end{equation}
where $\hbar k_{x}=p_{x}$ and $t$ is the duration of the Raman pulse. The spread of $k_{x}$ in the DFG causes damping of the total spin dynamics due to Doppler shifts. By integrating Eq.~(\ref{osc_eq}) over the measured momentum distribution of the DFG, the Rabi spin dynamics of a non-interacting DFG are obtained as plotted in Fig.~\ref{rabifig}(a).

However, elastic collisions can change an atom's momentum along the SOC dimension $\hat{x}$, which  modifies the spin dynamics and results in additional Rabi-oscillation dephasing.  We now describe this process by referring to the illustration in Fig.~\ref{rabifig}(b).  The atoms begin on the $\ket{\downarrow}$ dispersion but project onto a superposition of the two dressed dispersions when the Raman-coupling diabatically turns on.  We take as examples the two atoms shown as green shaded ovals at the center of the figure.  Each oval covers  green circles that  depict the $\ket{\downarrow}$-components of the dressed state belonging to a single atom. The size of each circle is proportional to the $t=0$ projection of dressed state onto the bare $\ket{\downarrow}$ state at the given quasimomentum. The atoms then begin to Rabi oscillate between $\ket{\downarrow}$ and $\ket{\uparrow}$, which in the figure would appear as exchanges in the sizes of the circles in each shaded oval over time. Only at the degeneracy point $p_{x}=0\hbar k_{R}$ is the Raman coupling resonant, yielding Rabi oscillations that exhibit full contrast.   Atoms at all other quasimomenta have an effective detuning (i.e., a Doppler-shift) that reduces the Rabi oscillation contrast, as is the case for the data in Fig.~\ref{rabifig}(a).  

Elastic collisions between two atoms can change both atoms' momenta along the SOC axis, which changes the effective Rabi detunings of each atom. For example, atoms represented by  hollow green circles could elastically scatter to states represented by solid green circles while conserving total momentum and energy (possibly by exchanging momentum along other dimensions). The subsequent spin dynamics depart from the pre-collision Rabi nutation, resulting in dephasing and a higher steady-state occupation of $\ket{\uparrow}$ not accounted for in  Eq.~(\ref{osc_eq}) describing non-interacting Fermi gases.

Collisional effects were negligible in previous SOC fermionic alkali gases, even in the presence of a sizeable $s$-wave scattering length of $169a_0$ between the spin states of $^{40}$K~\cite{Wang2012,Cheuk2012}. However, elastic dipolar collisions due to $^{161}$Dy's large dipole moment (10 Bohr magnetons) are not negligible on the timescale of $\Omega_{R}^{-1}$ in our system. In order to account for momentum changing collisions, we use a Monte Carlo simulation that includes elastic collisions between pairs of atoms and is described in Appendix~\ref{elasticDDI}.  When fitting the data, the rate of collisions and Raman coupling strength are treated as free parameters. Though the simple Monte Carlo model does not take into account correlations involving the atoms' positions or other details of the collisions, the model does capture the increased damping of the oscillations and higher steady state of fraction of $\ket{\uparrow}$ atoms.

Curves from both the Monte Carlo simulation and the non-interacting case of Eq.~(\ref{osc_eq}) are shown with data in Fig.~\ref{rabifig}(a). The fitted elastic collision rate of 9.9(2)~$\mu$s$^{-1}$ is larger than the expected rate from dipolar collisions, 3(1)~$\mu$s$^{-1}$~\footnote{We note that the particular fitted value of the collision rate has negligible effect on the fitted value of $\Omega_{R} = 6.57(4)E_{R}$; the covariance is small, $7\times10^{-4}$.}. However, a more complete simulation, perhaps using the direct simulation Monte Carlo technique (DSMC) of Ref.~\cite{Sykes2015}, may yield a more accurate collision rate by taking into account all correlations between position, spin, and momentum. Additional consideration will be given to any contribution from the as yet unmeasured $s$-wave scattering length between the $m_{F}=-21/2$ and $m_{F}=-19/2$ states; this is beyond the scope of this present work.  These measurements indicate the achievement of a weakly interacting Fermi gas under the influence of the Raman-coupling necessary to generate synthetic gauge fields, and complements lattice-bound  fermionic polar molecule systems in which dipolar interactions can also influence spin dynamics~\cite{Yan:2013fn}.

\emph{Inelastic dipolar decay of the SOC dressed state.} In addition to elastic collisions, the DDI also induces two-body inelastic collisions (dipolar relaxation), which can cause heating or atom loss. When the atoms are adiabatically loaded into the SOC dressed states of Eq.~(\ref{SOH1}), the spinor is a superposition of the bare spin states. The $\ket{\uparrow}$ portion of an atom's spin-state can undergo dipolar relaxation even when in the lowest dressed band shown in blue in Fig.~\ref{socfig}(b). This is especially significant if the Zeeman energy is larger than twice the trap depth because a single spin flip releases sufficient energy to eject both atoms from the trap, as is the case for fields above $\sim$50 mG in our optical dipole trap.

\begin{figure}[t!]
  \includegraphics[width=\columnwidth]{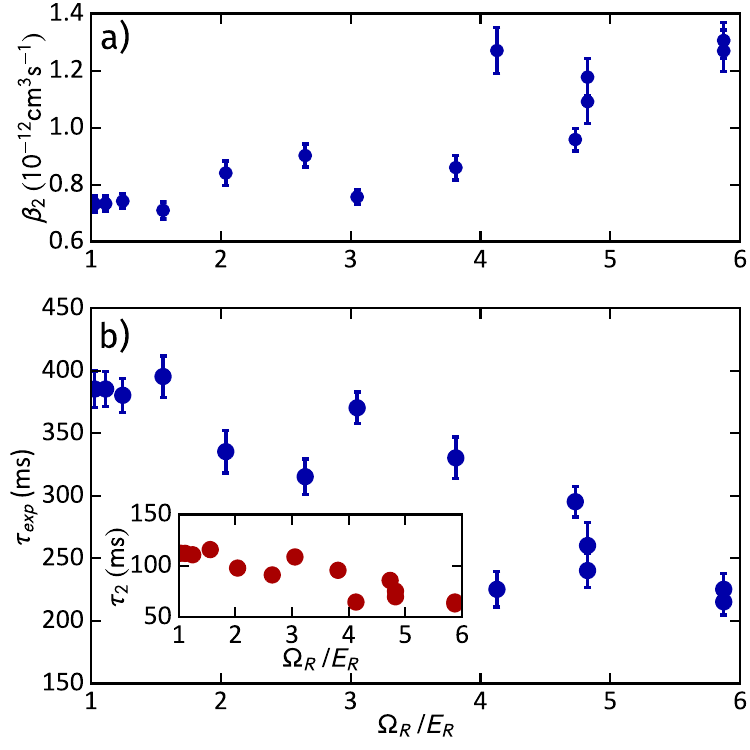}
  \caption{(a) Two-body loss parameters for $\delta=0.0(3)E_{R}$.
    (b) Experimental lifetimes for a gas of $1\times10^{4}$ $^{161}$Dy atoms with $T/T_{F}=0.4$ and $\bar{f}=150$~Hz, where the ratio of the dipolar  and Fermi energies fall by $e$ in time $\tau_{exp}$. (inset) Characteristic lifetimes $\tau_{2}=(\bar{n}(0)\beta_{2})^{-1}$ with  $\bar{n}(0)=1.2$$\times$$ 10^{13}$~cm$^{-3}$.}
  \label{betas}
\end{figure}

We measure the dipolar relaxation out of the SOC-dressed states by varying the hold time $\tau_{soc}$ in the adiabatically loaded dressed states. After turning off the coupling and releasing the atoms from the trap, we measure the atom number via absorption imaging and fit the atom loss curve to a numerically integrated rate equation.  The equation has terms for both one-body loss $\beta_{1}$ (fixed parameter) and two-body loss $\beta_{2}$ (free parameter):
\begin{equation}
\frac{dN}{dt} = -\beta_{1}N - \beta_{2}\bar{V}^{-1}N^{2},
\end{equation}
where $\bar{V}=\sqrt{8}(2\pi)^{3/2}\sigma_{x}\sigma_{y}\sigma_{z}$ is the mean collisional volume for a harmonically trapped thermal gas of Gaussian width $\sigma_{i}$. We fix $\beta_{1}$ to the background scattering rate (20~s)$^{-1}$.  No heating due to the Raman lasers is measured, confirming our expectation that this $L=6$ atom would be immune to spontaneous emission under SOC~\cite{Cui2013}. The fitted values of $\beta_{2}$ for different coupling strengths and $\delta=0.0(3)E_{R}$ are shown in Fig.~\ref{betas}(a).  These may be compared with the dipolar relaxation rates found in Ref.~\cite{Burdick2015} for non-SOC Dy gases at 1~G:  $[0.3,10,30]\times10^{-12}$ cm$^3$s$^{-1}$ for identical fermions, distinguishable particles, and identical bosons, respectively.

Because two-body loss is density dependent, characteristic timescale $\tau_{2}$ can only be found by multiplying  the loss parameter by a  density:  $\tau_{2} = (\bar{n}(0)\beta_{2})^{-1}$, where $\bar{n}(0)=N/\bar{V}$ is a typical initial density for our system. Alternatively, since one may care most about preserving interactions in the system, one can define the relevant experimental timescale as the time required for the ratio of the dipolar energy to Fermi energy $\epsilon(t) = E_{dd}(t)/E_{F}(t)$ to decrease to $\epsilon(\tau_{exp})=\epsilon(t=0)/e$, where $E_{dd}=N \mu_{0} \mu^{2}(48\pi^{3/2}\sigma_{x}\sigma_{y}\sigma_{z})^{-1}$ \cite{Sykes2015} and $E_{F}=h \bar{f}(6N)^{1/3}$. The values of $\tau_{exp}$ and $\tau_{2}$ shown in Fig.~\ref{betas}(b) are for a gas of $N=1\times10^{4}$ atoms at $T/T_{F}=0.4$ and $\bar{f}=150$~Hz. For such a gas, $\epsilon(0)=0.56\%$ and $\bar{n}(0)=1.2\times10^{13}$~cm$^{-3}$. 

The dipolar relaxation cross-sections in free space (no SOC) have been calculated under the first-order Born approximation~\cite{Hensler2003,Pasquiou2010} and have been verified for Bose, Fermi, and distinguishable statistics \cite{Burdick2015} in free-space (i.e., in the absence of SOC) for fields below 1.3~G. Notably, the rate of inelastic collisions depends on both the relevant quantum statistics as well as on the amount of energy released. Of particular interest, dipolar relaxation is suppressed for identical fermions as the magnetic field is increased: $\beta_{2}\propto 1/\sqrt{B}$ \cite{Burdick2015,Pasquiou2010}.

To test the importance of fermionic suppression in achieving long lifetimes, we also measured the $^{161}$Dy lifetime at low-field 0.48(1)~G with $\Omega_{R}=5.9(3)E_{R}$ and $\delta=-0.1(2)$. At this low magnetic field, the quadratic Zeeman shift is insufficient to limit Raman coupling to only the $\ket{m_{F}=-21/2}$ and $\ket{m_{F}=-19/2}$ states, so most collisions involve many $m_{F}$ states and are distinguishable. Furthermore, any indistinguishable collisions have less suppression due to the lower Zeeman energy released. Indeed, the measured low-field two-body loss parameter at this field is $\beta_{2}=2.1(4)\times10^{-11}$~cm$^3$/s.  This is $\sim$16$\times$ larger than the high-field two-body loss parameter $\beta_{2}=1.28(2)\times10^{-12}$~cm$^3$/s for the same $\Omega_{R}$ at $B_0$. The corresponding low-field characteristic timescale at density $\bar{n}=1.2\times10^{13}$~cm$^{-3}$ is only $\tau_{2}=4.0(8)$~ms.  We also measured similarly short, $<$10-ms lifetimes  for Raman-dressed bosonic Dy, even at low field where $\beta_{2}$ should be minimized for distinguisable particles and Bosons.  These results imply that proposals for observing exotic non-Abelian spinors and many-body physics using SOC dipolar BECs will be quite difficult to realize~\cite{Lian:2012eo,*Deng:2012kh,*Gopalakrishnan:2013cb,*Cui2013,*Lian:2014dt}. 

  \begin{figure}[t!]
  \includegraphics[width=\columnwidth]{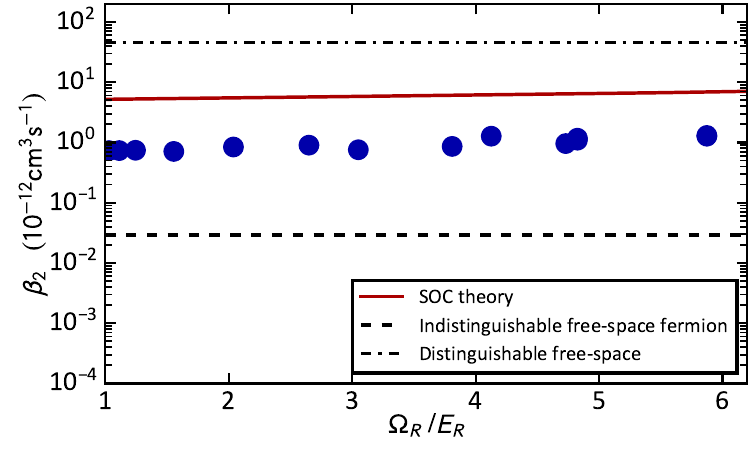}
  \caption{Measured and theoretical two-body loss parameters at $B_0$ with $\delta=0.0(3)E_{R}$. The black dashed (dot-dashed) line is calculated for a gas of identical dipolar fermions (distinguishable particles) with spin mixtures under no SOC. The solid red line is calculated for SOC dipolar fermions under using Eq.~(\ref{beta_integral}).}
  \label{betatheory}
\end{figure}

\textit{Comparison of loss rate to theory.}  Before we describe a theory developed to  predict the loss rates of SOC degenerate dipolar Fermi gases, we present a comparison between our data and the results of this theory.  Figure~\ref{betatheory} plots three theory curves along with the data of Fig.~\ref{betas}(a).  The loss rates for distinguishable Dy-like atoms and indistinguishable Dy fermions with spin mixtures in the absence of SOC are shown along with the result from our SOC theory for fermions. Distinguishable (indistinguishable) non-SOC theory overestimates (underestimates) the collision rate by a factor of $\sim$40. The scaling of $\beta_{2}$ with $\Omega_{R}$ is captured by the SOC theory, but the theory predicts a collision rate $\sim$6$\times$ larger than the measured rate. Because our calculation, based on Eq.~(\ref{beta_integral}) below, involves only a simple  average over the  ensemble of atoms, it discards all correlations between position and momentum. A simulation of the dynamics with the SOC gas using DSMC, or a similar technique, may yield a more accurate collision rate.  We note that this theory does not incorporate loss from Feshbach resonances, yet still predicts a loss rate greater than our data, implying that such loss may not play a prominent role in determining the stability of these SOC gases at high field.

\textit{Description of SOC dipolar loss calculation.}
We now describe our approximate calculation of dipolar relaxation rates in the SOC dipolar Fermi gas.  We do so in the SOC rotating frame, using an approach that is similar to that found in Refs.~\cite{Zhang2013,*Zhang2013a} for calculating loss from metastable dressed bands in a non-dipolar gas.  We consider scattering from the two-atom state 
$\ket{\gamma}=\ket{\mathbf{p}_{1}}\ket{\alpha_{1}(\mathbf{p}_{1})}\ket{\mathbf{p}_{2}}\ket{\alpha_{2}(\mathbf{p}_{2})}$, where $\ket{\alpha_{i}(\mathbf{p}_{i})}$ is the spinor in the dressed band $\alpha_{i}$ and $\mathbf{p}_{i}$ is the quasimomentum of atom $i$,
to the final state $\ket{\gamma^{\prime}}$, which exists in a lower dressed-state manifold $\mathcal{M}$~\cite{cohen1998atom} labeled by the imbalance $M$ between the number of photons in one Raman-beam versus the other.
We aim to calculate the two-body parameter
\begin{equation}
\begin{aligned}
\beta_{2}=\frac{8}{m^{2}}\sum_{\alpha^{\prime}_{1},\alpha^{\prime}_{2}}
\int d\mathbf{p}_{1}& d\mathbf{p}_{2} d\mathbf{p}^{\prime}_{1}d\mathbf{p}^{\prime}_{2}
\delta_{E}\delta_{\mathbf{p}^{\prime}} \\
& \times \Pi(\mathbf{p}_{1})\Pi(\mathbf{p}_{2})
\left| f(\gamma^{\prime},\gamma)\right|^{2},
\label{beta_integral}
\end{aligned}
\end{equation}
where $f(\gamma^{\prime},\gamma)$ is the scattering amplitude between $\gamma$ and $\gamma^{\prime}$; $\Pi(\mathbf{p}_{i})$ is the initial quasimomentum distribution of particle $i$; and $\delta_{E}$ and $\delta_{\mathbf{p}^{\prime}}$ ensure conservation of energy and momentum, respectively. The summation over $\alpha_{i}$ includes all dressed bands, and we assume the atoms are initially in the lowest band.

Because the gas is adiabatically loaded into the SOC-dressed band, the quasimomentum distribution is simply the Fermi-Dirac momentum distribution of the initial gas
\begin{equation}
\Pi(\mathbf{p}_{i}) = \frac{\mathrm{Li}_{3/2}[-\zeta e^{-p_{i}^{2}/(2m k_{B}T)}]}{\left( 2\pi m k_{B}T\right)^{3/2}\mathrm{Li}_{3}[-\zeta]},
\end{equation}
where $\mathrm{Li}_{s}[x]$ is the polylogarithm series and $\zeta$ is the fugacity of the DFG.

All that remains is to determine the dipolar relaxation scattering amplitude in the SOC frame, $f(\gamma^{\prime},\gamma)$. We first recall the transformations necessary to produce the Hamiltonian in Eq.~(\ref{SOH1}). In the lab frame, the single particle SOC Hamiltonian for particle $i$ is
\begin{equation}
  \begin{aligned}
    \hat{H}_{i}(t) = \frac{\hat{\mathbf{p}}_{i}^{2}}{2m} - &\hbar\omega_{Z}\hat{F}_{z,i} \\&- \frac{\hbar\Omega_{R}}{2}\left(e^{i(2k_{R}x_{i}-\Delta t)}\hat{\Lambda}_{i}+\mathrm{H.c.}\right).
    \end{aligned}
\end{equation}
We apply two transformations $\hat{U}_{t,i}=e^{-i\Delta t\hat{F}_{z,i}}$ and $\hat{U}_{k,i}=e^{i2k_{R}x_{i}\hat{F}_{z,i}}$ to recover the form of Eq.~(\ref{SOH1}) with unmodified parabolic dispersions along $y$ and $z$. The SOC Hamiltonian for atom $i$ is then
\begin{equation}
\begin{aligned}
\hat{H}_{i} =  \hat{H}_{R,i} + \frac{\hat{p}_{y,i}^{2}+\hat{p}_{z,i}^{2}}{2m}.
\end{aligned}
\end{equation}
This is extended to two atoms, as required for dipolar relaxation:
\begin{equation}
  \hat{H}_{12} = \hat{H}_{1}\otimes\hat{\mathbb{I}}_{2} + \hat{\mathbb{I}}_{1}\otimes \hat{H}_{2},
  \label{SOH2}
\end{equation}
where $\hat{\mathbb{I}}_{i}$ is the identity in the subspace of atom $i$.

The dipole-dipole interaction has the form
\begin{equation}
  \hat{H}_{dd} = \frac{\mu_{0}(g_{F}\mu_{B})^{2}}{4\pi r^{3}}
\left[ \hat{\mathbf{F}}_{1} \cdot \hat{\mathbf{F}}_{2} - 3(\hat{\mathbf{F}}_{1}\cdot \tilde{\mathbf{r}}) (\mathbf{F}_{2}\cdot \tilde{\mathbf{r}}) \right],
\label{DDIH}
\end{equation}
where $\mathbf{r}=\mathbf{r}_{1}-\mathbf{r}_{2}$ and $\tilde{\mathbf{r}}=\mathbf{r}/|\mathbf{r}|$. Combining Eq.~(\ref{SOH2}) and (\ref{DDIH}) into the full Hamiltonian yields
\begin{equation}
\hat{H}_{\phantom{dd}} = \hat{H}_{12} + \hat{H}_{dd}^{\prime},
\label{FullH}
\end{equation}
where 
\begin{equation}
  \hat{H}_{dd}^{\prime} = \hat{U}^{\dagger}_{k,1}\hat{U}^{\dagger}_{k,2}
  \hat{U}^{\dagger}_{t,2}\hat{U}^{\dagger}_{t,1}\hat{H}_{dd}\hat{U}_{t,2}\hat{U}_{t,1}
  \hat{U}_{k,1}\hat{U}_{k,2}.
\end{equation}
Although both single- and double-spin-flip inelastic processes can occur, we restrict our consideration to single-flip events. (Double-flip processes for $\ket{\uparrow}$ are slower by roughly a factor of $F=21/2$ \cite{Burdick2015}.) The single-spin-flip portion is
\begin{equation}
  \hat{H}_{dd}^{(1)}\propto \left[\hat{F}_{z,1}\hat{F}_{-,2}+\hat{F}_{-,1}\hat{F}_{z,2}\right],
\end{equation}
which in the SOC frame becomes
\begin{equation}
\hat{H}_{dd}^{\prime(1)}\propto e^{i\Delta t}\left[e^{-i2k_{R}x_{2}}\hat{F}_{z,1}\hat{F}_{-,2}+e^{-i2k_{R}x_{1}}\hat{F}_{-,1}\hat{F}_{z,2}\right].
\label{Flip1SOC}
\end{equation}

    \begin{figure}[t]
    \includegraphics[width=\columnwidth]{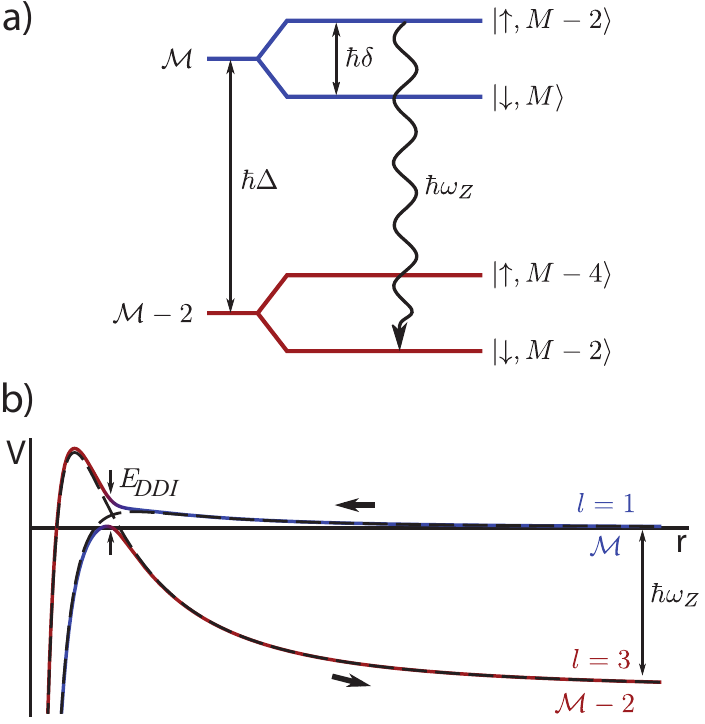}
    \caption{(a) A system initially in photon number imbalance manifold $\mathcal{M}$ can decay to manifold $\mathcal{M}-2$ via the DDI, releasing the Zeeman energy $\hbar\omega_{Z}$ as kinetic energy. Each state is labeled by the atom's spin state  and the Raman lasers' photon number imbalance. The DDI can only couple  $\ket{\uparrow}$ to $\ket{\downarrow}$ and cannot change the photon imbalance. (b) Interatomic potentials for identical fermions entering a collision on the blue $p$-wave ($l=1$) orbital potential and exiting on the red $f$-wave ($l=3$) orbital potential. The potentials are shown in the rotating SOC frame and with $\omega_Z\gg\delta$. The DDI couples these two potentials with energy $E_{DDI}$.  Atoms may exit on the lower dressed potential, releasing  Zeeman energy $\hbar\omega_{Z}$. The uncoupled potentials are shown as black dashed lines.}
    \label{cartoons}
    \end{figure}

With this Hamiltonian, the single-spin flip scattering amplitude under the first-order Born approximation in the SOC frame is
\begin{widetext}
\begin{equation}
\begin{aligned}
  f^{(-1)}(\gamma^{\prime},\gamma) = 
  \frac{m\mu_{0}(g_{F}\mu_{B})^{2}}{16\pi\hbar^{2}}\sqrt{\frac{24}{5\pi}}
  e^{i\Delta t}\bigg\langle \gamma^{\prime}\bigg|\frac{Y^{*}_{2,-1}(\mathbf{\hat{r}})}{r^{3}} 
  \bigg[e^{-i2\hbar k_{R}x_2}\hat{F}_{z,1}\hat{F}_{-,2}
      +e^{-i2\hbar k_{R}x_1}\hat{F}_{-,1}\hat{F}_{z,2}\bigg]
\bigg|\gamma\bigg\rangle,
\end{aligned}
\label{scatteringamplitude}
\end{equation}
\end{widetext}
where $Y_{lm}(\mathbf{\hat{r}})$ is a spherical harmonic. We limit ourselves to the first-order Born approximation because our reproduction of the calculation in Ref.~\cite{Zhang2013} for metastable dressed-band decay showed that higher order corrections in the Lippman-Schwinger equation only contributed at the $10^{-4}$ level. The modification of the inelastic DDI in the SOC rotating frame is due to the commutation relation of  $\hat{F}_{-}$ and $\hat{F}_{z}$. The time dependent exponential does not affect the calculation of Eq.~(\ref{beta_integral}) because Eq.~(\ref{scatteringamplitude}) only enters as the modulus squared. However, the position dependent exponentials change the quasimomentum conservation condition: Whereas conservation of momentum would typically require $\mathbf{p}_{1}+\mathbf{p}_{2}=\mathbf{p}_{1}^{\prime}+\mathbf{p}_{2}^{\prime}$,  the quasimomentum conservation condition is instead $\mathbf{p}_{1}+\mathbf{p}_{2}=\mathbf{p}_{1}^{\prime}+\mathbf{p}_{2}^{\prime} + 2\hbar k_{R}\hat{x}$. Thus, quasimomentum is conserved up to $2\hbar k_{R}$ for single-spin-flip dipolar relaxation in the SOC frame, which is due to the atom changing spin without exchanging a photon between the Raman laser fields and thereby coupling to a dressed-state manifold of differing Raman-beam photon number imbalance.

An intuitive picture for the relaxation events may be formed  in which the system decays, upon dipolar relaxation, from photon imbalance manifold $\mathcal{M}$ to manifold $\mathcal{M}-2$, as illustrated in Fig.~\ref{cartoons}(a). Within each manifold the system state is labeled by the atom's spin state and the Raman lasers' photon imbalance. For an atom initially in $\ket{\downarrow}$ with photon imbalance $M$, the state can be written as
\begin{equation}
  \ket{\Phi_{\mathcal{M}}}=\phi_{\downarrow}\ket{\downarrow,M} + \phi_{\uparrow}\ket{\uparrow,M-2},
  \label{initialstate}
\end{equation}
because the atom's spin state is coherently changed by exchanging a photon from one laser to the other, changing the imbalance by 2. By contrast, if the DDI incoherently changes the $\ket{\uparrow}$ portion of Eq.~(\ref{initialstate}) to $\ket{\downarrow}$, the atom's new dressed state can be written as
\begin{equation}
  \ket{\Phi_{\mathcal{M}-2}^{\prime}}=\phi_{\downarrow}^{\prime}\ket{\downarrow,M-2} + \phi_{\uparrow}^{\prime}\ket{\uparrow,M-4}.
  \label{finalstate}
\end{equation}
The coefficients $\phi_{\sigma}$ ($\phi_{\sigma}^{\prime}$) depend on the initial (final) dressed band and momentum. The transition rate between Eq.~(\ref{initialstate}) and (\ref{finalstate}) then depends on their overlap under dipolar relaxation,
\begin{equation}
  \bra{\Phi_{\mathcal{M}-2}^{\prime}}H_{dd}\ket{\Phi_{\mathcal{M}}} = \phi_{\uparrow}\phi_{\downarrow}^{\prime}\bra{\downarrow,M-2}H_{dd}\ket{\uparrow,M-2},
\end{equation} which results in a coupling energy $E_{DDI}$.
Whereas the energy required to move a photon from one laser to the other compensates for the Zeeman energy difference (up to an amount $\delta$) in a coherent process, the lack of photon exchange in dipolar relaxation, concomitant with the flipped spin, causes the full Zeeman energy to be released into the system as kinetic energy.

The relaxation process may also  be viewed as a transition between two  molecular potentials~\cite{Beaufils2008}. As illustrated in Fig.~\ref{cartoons}(b), identical fermions colliding on the $p$-wave ($l=1$)  potential can exit on an $f$-wave ($l=3$)  potential of different photon imbalance number $\mathcal{M}$, with a transition rate set by the DDI coupling.  Fermionic suppression of dipolar relaxation occurs because inelastic dipolar collisions, unlike elastic dipolar collisions, are a short-range process~\cite{Pasquiou2010,Burdick2015}: identical ultracold fermions must surmount the $p$-wave centrifugal barrier to collide, leading to a kinematic suppression.

\textit{Conclusion.} While creating spin-orbit coupled fermionic alkali atoms using Raman coupling has been very successful, the short spontaneous-emission-limited lifetime of alkali atoms severely hampers investigations of interacting topologically non-trivial systems.  We have demonstrated that spin-orbit-coupling in lanthanide atomic gases, specifically $^{161}$Dy, can have much longer lifetimes that are instead limited by inelastic dipolar collisions. Fortunately, the dipolar relaxation is not rapid, allowing the study of spin-orbit coupled degenerate dipolar Fermi gases of Dy  for as long as 400 ms, far longer than in fermionic alkali gases.    Looking forward, the elastic dipolar collisions observed here may be exploited to study interacting spin-orbit coupled Fermi gases.  Moreover, the lifetime of these Raman-coupled dipolar Fermi gases may now be sufficiently long to create and study exotic quantum liquids~\cite{Nascimbene:2013fr} as well as quantum Hall ribbons in synthetic dimensions~\cite{Celi:2014dg,Mancini:2015fba,Stuhl:2015cb} where the large Dy spin provides a wide bulk with which to isolate edge states. 

We thank J.~Bohn, W. Kao, J.~DiSciacca, A.~Papageorge, S.~Gopalakrisnan and B.~Lian and acknowledge support from the AFOSR and NSF.

\appendix

\section{Monte-Carlo simulation of the collisional contribution to Rabi oscillations}
\label{elasticDDI}
To produce  the theory curve in Fig.~\ref{rabifig}(a), each simulated atom is assigned an initial momentum sampled from the Fermi-Dirac momentum distribution and has a spinor of the form
\begin{equation}
  \ket{\psi(t)}=c_{\uparrow}(t)\ket{\uparrow}+c_{\downarrow}(t)\ket{\downarrow},
\end{equation}
initialized to  $c_{\downarrow}(0)=1$ and $c_{\uparrow}(0)=0$. The spinor of each atom then evolves according to
\begin{widetext}
  \begin{equation}\label{elasticDDIrabi}
      c_{\downarrow,\uparrow}(t) = \left\{c_{\downarrow,\uparrow}(0)\cos(t\Theta/2)\pm i\left[\frac{\delta}{\Theta}c_{\downarrow,\uparrow}(0)\pm\frac{\Omega_{R}}{\Theta}c_{\uparrow,\downarrow}(0)+\frac{2\hbar}{m\Theta}k_{R}(k_{R}+k_{x})c_{\downarrow,\uparrow}(0)\right]\sin(t\Theta/2)\right\}e^{-i\omega t},
  \end{equation}
  \begin{equation}
    \omega=\frac{\hbar(2k_{R}^{2}+2k_{R}k_{x}+k_{x}^{2})}{2m}, \qquad
    \text{and}\qquad\Theta = \sqrt{4\left[\frac{\hbar k_{R}(k_{R}+k_{x})}{m}\right]^{2}+4\delta \left[\frac{\hbar k_{R}(k_{R}+k_{x})}{m}\right] +\delta^{2}+\Omega_{R}^{2}}.\nonumber
  \end{equation}
\end{widetext}
Simulated collisions between atoms occur with a rate that is free to be adjusted to fit the data.  The Raman coupling rate $\Omega_R$ is also a free parameter.  Averages over all simulated atomic spinors produce the solid-line Rabi-oscillation curve in Fig.~\ref{rabifig}(a). 

For two atoms colliding at time $t=t^{\prime}$, the spinor of each atom does not change immediately, i.e., $c_{\downarrow}(t^{\prime-})=c_{\downarrow}(t^{\prime+})$ and $c_{\uparrow}(t^{\prime-})=c_{\uparrow}(t^{\prime+})$ for both atoms. However, the collision does change the momentum of each particle, i.e., $k_{x}(t^{\prime-})\neq k_{x}(t^{\prime+})$, and thus changes the subsequent evolution of the spinors through Eq.~(\ref{elasticDDIrabi}). This results in not only more dephasing but also an equal steady-state population of $\ket{\uparrow}$ and $\ket{\downarrow}$ because the collisions effectively erase the effect of the initial conditions $c_{\downarrow}(0)=1$ and $c_{\uparrow}(0)=0$.

%

\end{document}